\documentclass{article}
\usepackage{graphicx} 
\usepackage[authoryear]{natbib}
\usepackage{enumitem}
\usepackage{amsmath}
\usepackage{booktabs, caption, longtable, colortbl, array}
\usepackage{makecell}
\usepackage{hyperref}
\usepackage{lineno}        
\usepackage{setspace}      
\usepackage[margin=1in]{geometry} 
\usepackage{authblk}

\begin{document}

\title{Estimating high-resolution albedo for urban applications}

\author[1,*]{David Fork}
\author[2,*]{Elizabeth Jane Wesley}
\author[1]{Salil Banerjee}
\author[1]{Vishal Batchu}
\author[1]{Aniruddh Chennapragada}
\author[1]{Kevin Crossan}
\author[1]{Bryce Cronkite-Ratcliff}
\author[1]{Ellie Delich}
\author[3]{Tristan Goulden}
\author[1]{Mansi Kansal}
\author[1]{Jonas Kemp}
\author[2]{Eric Mackres}
\author[1]{Yael Mayer}
\author[1]{Becca Milman}
\author[1]{John C. Platt}
\author[1]{Shruthi Prabhakara}
\author[1]{Gautam Prasad}
\author[1]{Shravya Shetty}
\author[1]{Charlotte Stanton}
\author[1]{Wayne Sun}
\author[1,4]{Lucy R. Hutyra}

\affil[1]{Google}
\affil[2]{World Resources Institute}
\affil[3]{National Ecological Observatory Network, Battelle, Boulder, Colorado, USA}
\affil[4]{Boston University}
\affil[*]{Co-Lead Authors}
\affil[ ]{\textbf{Corresponding Author:} David Fork (fork@google.com)}

\date{July 2025}

\maketitle

\begin{abstract}

Implementation of cool roofs is a high-impact pathway for mitigating heat at both global and city scales. However, while albedo estimates derived from Sentinel-2 are free and globally-available, the 10 m resolution is insufficient to resolve individual roofs. We present methods for increasing the resolution of Sentinel-2 albedo using high-resolution satellite imagery to produce albedo inferences at a 30-cm scale. Validating against high-resolution aerial albedo measurements over Boulder, CO we find improved precision and accuracy relative to Sentinel-2 with an RMSE of 0.04. Applying these methods to 12 global cities, we evaluate the impacts of three cool roof implementation scenarios. We find that cities can see up to a 0.5°C cooling effect from full scale implementation of cool roofs and prioritizing the largest buildings for implementation is a highly effective policy pathway. While Sentinel-2 produces accurate estimates of albedo change at larger scales, high-resolution inferences are required for prioritizing buildings based on their solar radiation management potential. This research demonstrates a scalable implementation of targeted cool roof interventions in neighborhoods with the greatest potential for heat mitigation by enabling actionable, building-level insights.

\end{abstract}

\section{Introduction}

By 2050 the global urban population is expected to increase from 55\% to 68\% \cite{unitednations2018}, a projected doubling of the global urban footprint \cite{huang2019, seto2012}. This projected urban land expansion is expected to cause warming almost as strong as that caused by global greenhouse gas emissions \cite{georgescu2014, huang2019}. Urbanization produces profound changes in the surface-energy balance through the conversion of vegetation to impervious surfaces, a change that typically reduces surface reflectivity—also known as albedo. At the global level, these changes increase radiative forcing, producing a small but significant global temperature increase \cite{ouyang2022}. At the local level, these land conversions cause a suite of climate modifications known as the urban heat island (UHI) effect, which causes urban air temperatures to elevate above background climate conditions \cite{oke1982}. These global and local effects of urbanization combined with overall global warming are increasing temperatures in cities at the same time that urban population growth is increasing exposure \cite{vahmani2019}. In fact, between 1950 and 2015 about 60\% of urban residents experienced warming twice the global average \cite{estrada2017}. High temperatures in cities are detrimental to human health and well-being, can damage infrastructure and disrupt services, and increase energy demand for air conditioning  \cite{santamouris2015}.

There are two primary city-scale pathways for lowering air temperatures in urban areas that do not rely on the use of air conditioning: the evaporation of water---primarily by vegetation---and the reflection of sunlight. Common strategies for these pathways include increasing tree canopy cover and other vegetation, and increasing the reflectivity of roofs and roads, respectively. Regional-scale simulations using a combination of these strategies suggest that air temperature reductions on the order of several degrees Celsius are feasible \cite{taha2021} and could have substantial positive health impacts \cite{vahmani2019}. 

Roofs occupy anywhere from 20-25\% of urban land areas \cite{akbari2008a} and vary greatly in size. The majority of urban roofs are covered with dark materials, absorbing as much as 80-90\% of incoming solar radiation \cite{santamouris2014}. Cool roofs are capable of reflecting almost all incident sunlight \cite{ban-weiss2015}. Increasing the reflectivity of roofs reduces the surface heat gain and the subsequent transfer of heat from the roof surface to both the atmosphere and the interior of the building, thereby lowering air temperatures and reducing energy loads \cite{akbari2009, ban-weiss2015}. While the average albedo of most cities ranges from 0.12–0.16 \cite{wu2024}---meaning that they reflect only 12-16 percent of received solar radiation---cool roofing materials can have initial solar reflectivities of above 0.9 (https://coolroofs.org/directory/roof) presenting a substantial opportunity for increasing city-wide albedos.

Cool roofs are a high-impact solution for mitigating urban heat that are relatively low cost compared to traditional roofing materials and produce substantial energy savings without considerable aesthetic conflict \cite{akbari2009}. Additionally, most roofs require regular maintenance necessitating periodic resurfacing, yielding ample opportunities for conversion to cool roofs \cite{akbari2009}. On the global level, conservative estimates show that increasing urban albedo by 0.1 would result in cooling equivalent to the removal of ~44 Gt of CO2 emissions \cite{akbari2009}. Locally, increasing the reflectivity of dark roofs can lower energy use by up to 20\% \cite{ban-weiss2015} and temperatures by up to 0.6°C per 0.10 albedo increase \cite{krayenhoff2021}. \cite{ouyang2022} suggests that highly reflective materials are not widely implemented at the global scale in current forms of urban land use, however, at both of these scales, a major source of uncertainty in modeling the effects of increasing albedo is estimating the modifiable land area \cite{akbari2012}.   

Accurate estimation of the potential for cool roof implementation and the temperature effects of wide-spread adoption requires a robust baseline estimation of rooftop albedo. This requires albedo imagery with a spatial resolution sufficiently high to resolve urban features relevant to cooling interventions, i.e. roofs. Further, the albedo inference should have an accuracy and revisit cadence that is useful for urban planning and progress monitoring. 

Accurate and globally available characterization of albedo can be derived from Sentinel-2 observations \cite{bonafoni2020}, however, the highest spatial resolution available is 10 meters. To effectively characterize their energy flux, features must be larger than the resolution of the data \cite{strahler1986}, and many urban roofs are well below the approximately 1000 $m^2$ size typically needed for reliable Sentinel-2–based albedo estimates. High resolution rooftop albedo inference from the National Agriculture Imagery Program (NAIP) multispectral (RGB + infrared) aerial images was introduced by \cite{ban-weiss2015} using atmospheric corrections and methods to radiometrically calibrate the NAIP cameras and more recently revisited by \cite{yi2025} who used machine learning methods and Sentinel-2 observations for calibration. Unfortunately, aerial imagery, such as NAIP, is not available globally, and the long acquisition times lead to the variation of illumination conditions.   

This paper presents simple, scalable, and validated methods for inferring high-resolution albedo by combining the spatial information from globally available high-resolution satellite imagery with the multi-spectral information from Sentinel-2, using Sentinel-2 as the source of radiometric calibration and atmospheric correction for the high-resolution imagery. To minimize illumination disparities (and to ensure global coverage of all habitable cities), we use sun synchronous high resolution satellite imagery which shares the roughly 10:30 AM local ascending node equatorial crossing time as Sentinel-2 (known as the “morning train or M-train). We quantify the accuracy of two methods for inferring high-resolution albedo—convolutional calibration and calibration transfer—and evaluate them along with Sentinel-2 albedo estimates against direct high-resolution aerial observations of albedo. Finally, we present an application of the methods, quantifying the solar radiation management potential for a cross-section of 12 global cities and estimate the albedo and temperature impacts of three city-scale cool roof implementation scenarios.  

\section{Results}

\subsection{NEON validation}
\subsubsection{Data}

\begin{figure}
    \centering
    \includegraphics[width=1\linewidth]{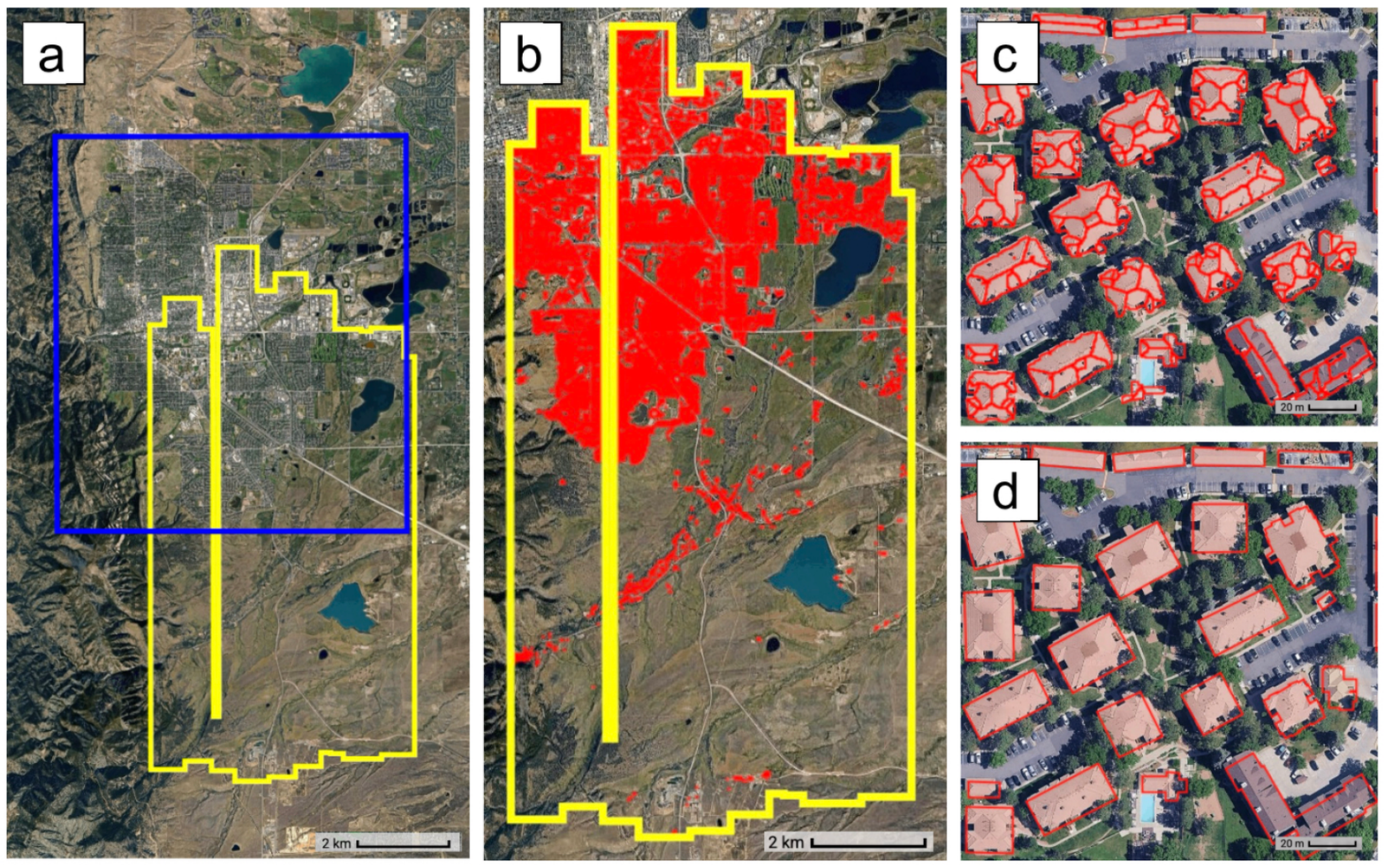}
    \caption{The boundaries of the NEON flightlines are shown in yellow, overlaid with satellite imagery to show the extent of the urban area of Boulder, CO. The blue in the left panel shows the extent of the coverage from the PNeo high-resolution imagery and the building footprints are shown in red in the middle panel. There is a gap in coverage towards the left where data is not available. The top right shows the roof segments and the bottom right the Overture Maps building footprints used in the analysis.
}
    \label{fig:maps}
\end{figure}

Direct ground-based albedo measurements (e.g., from flux towers or pyranometers) are sparse, making it difficult to validate large-scale satellite-derived albedo estimates, especially for urban areas. To address this challenge, we used albedo estimates produced by the National Ecological Observation Network (NEON) from hyperspectral aerial imaging flown over Boulder, CO on May 8, 2024 (Fig.~\ref{fig:maps}). The NEON albedo data has a resolution of 1 meter and is derived by integrating surface reflectance across wavelengths in the 0.4 to 2.5 micron band, weighted by the global flux received at the ground (https://data.neonscience.org/data-products/DP2.30011.001). Data are provided as 18 overlapping images from flightlines taken throughout the day, with solar elevations ranging from 42.65° to 61.45°.

To create the high-resolution albedo inference, we used two methods to radiometrically calibrate a high-resolution image with a low-resolution image—convolution calibration and calibration transfer. First, we created a temporal median of cloud and snow free low-resolution images proximate in time to a high resolution image. Then, we co-registered the two images and applied the calibration methods, described in detail below, to preserve the spatial structure of the high-resolution image while forcing it to agree radiometrically with the low-resolution image. See the Methods section for details.

For the high-resolution imagery, we made arrangements with Airbus in advance of the NEON flyovers to task the Pléiades Neo (PNeo) satellite constellation for dates on and near to the NEON flyover (May 8, 2024 and May 16 2024). We also analyzed PNeo imagery available for May 23, 2023. The Pléiades Neo constellation acquires imagery across six multispectral bands at a native 1.2-meter resolution and also captures a co-registered 30 cm panchromatic band. This configuration enables pansharpening, which fuses the two data types to create a final multispectral product with a high spatial resolution of 30 cm. 

For the source of the radiometric calibration, we created 100-day Sentinel-2 cloud and snow masked temporal medians centered on the PNeo date. Additionally, serendipity allowed the acquisition of a single, cloud-free Sentinel-2 image from April 30, 2024. While the temporal medians represent stable albedo estimates under changing cloud cover and ground conditions, we used the single calibration image to evaluate the methods while minimizing the impacts of these effects.

The validation study area is the intersection of the bounds of the PNeo and NEON datasets and covers much of the urbanized area of Boulder, CO, U.S.A. To compare the four datasets, we aggregated the albedo estimates to roof segments produced from digital surface models (DSMs) that were generated from each of the PNeo high-resolution images using the method of \cite{batchu2024}  (Fig.~\ref{fig:maps}). Because the roof segment polygons, which enclose areas of constant surface normal,  are derived directly from the PNeo imagery, each image date creates a slightly different set of roof segments that are self-aligned to the PNeo imagery. We calculated the median albedo within each roof segment for each albedo inference.  Having the surface normal for each roof segment polygon allows us to correct for differences in illumination due to directional reflectance by calculating the surface albedo from the image albedo as described in the Methods section.

Image albedo refers to the observed reflectance captured by the imaging sensor and is interpreted as if the scene were flat. However, reflectance varies with solar angle, view geometry, surface geometry, and the surface bidirectional reflectance distribution function (BRDF). Surface albedo represents the intrinsic reflectivity of the material under standardized illumination conditions. We corrected for these angular effects using the known orientation and tilt of roof planes with the assumption the surfaces are Lambertian, enabling a more accurate estimation of true surface albedo (see Methods for further details). While information about the roof geometry provides valuable context for evaluating our methods and may be accessed through the Google Maps Platform Solar API (https://developers.google.com/maps/documentation/solar), they are not publicly available. To validate our methods for a broader use case, we additionally calculated the median albedo of each dataset for building footprints acquired from OvertureMaps (Overture Maps Foundation, overturemaps.org).

To minimize the measurement inaccuracies resulting from differences in solar illumination, surface, and sensor viewing geometry, as well as atmospheric disturbances, we used only the highest precision NEON estimates, which we defined as roof segments covered by more than one flightline having a standard deviation of less than 0.01 between the multiple albedo measurements. The Kolmogorov-Smirnov (KS) test reports a KS statistic of $D=0.1554$ on a scale of 0 to 1, indicating that the distribution of the highest precision NEON observations is similar to the distribution of all NEON observations with a maximum distance of 15.54\% between the two cumulative distribution functions. We conclude that these highest precision estimates are representative of the entire NEON albedo dataset and define these observations as ground truth.

\subsubsection{Validation}

\begin{figure}
    \centering
    \includegraphics[width=1\linewidth]{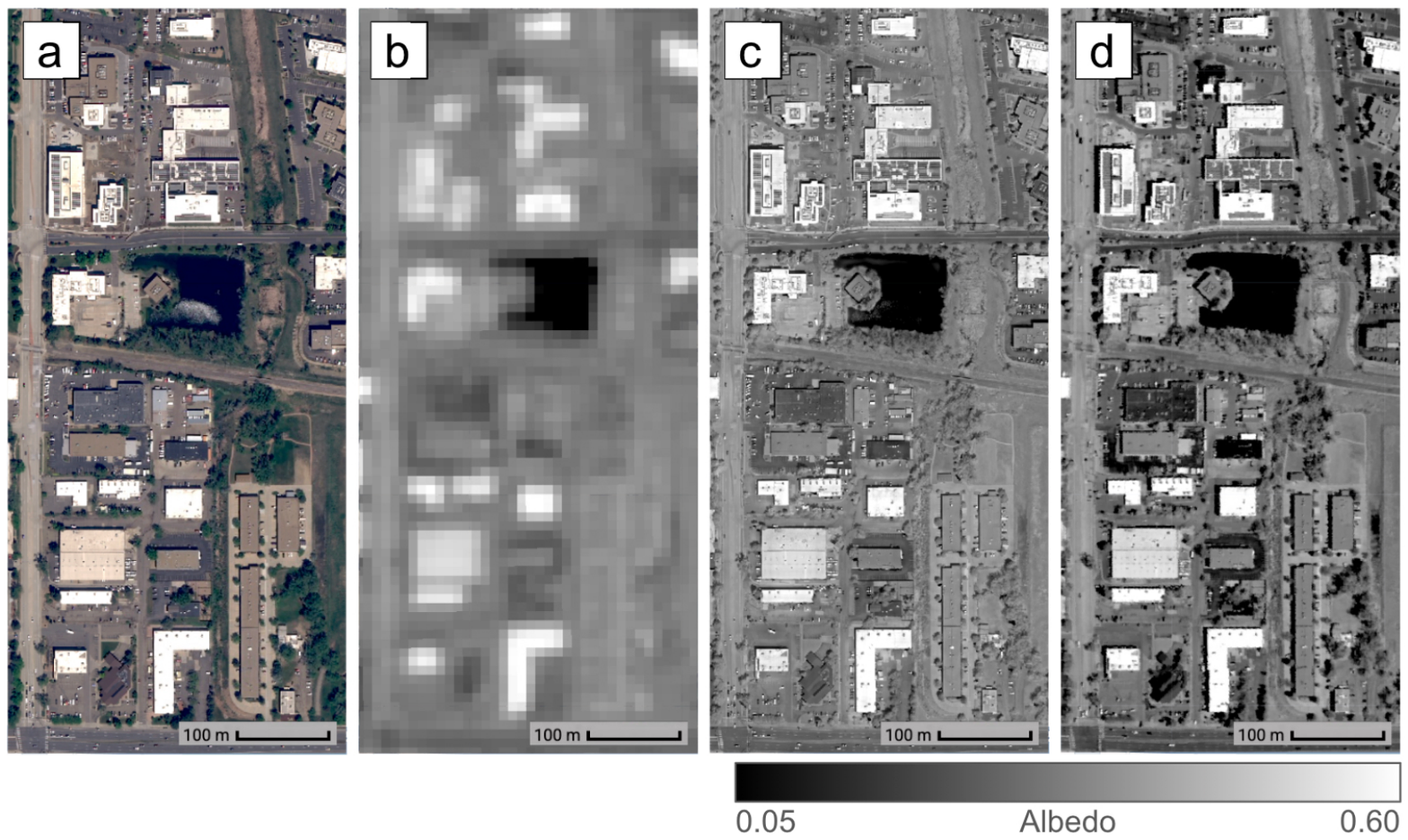}
    \caption{From left to right, (a) high resolution (30 cm) RGB imagery (Attribution: Pléiades Neo 3 and Pléiades Neo 4: © 2024 CNES / Airbus), (b) Sentinel-2 albedo (10 meters), (c) high-resolution albedo inference (30 cm), (d) NEON albedo (1 meter). Images are from the study area in Boulder, CO.
}
    \label{fig:resolution}
\end{figure}

To validate our methods, we compared four datasets (Fig.~\ref{fig:resolution}):

\begin{enumerate}
\item albedo measured at 1 meter resolution by the NEON hyperspectral instrument
\item albedo estimated at 10 meter resolution using Sentinel-2 imagery
\item albedo inferred by calibrating a 30 cm PNeo image using Sentinel-2 imagery using the convolutional calibration method 
\item albedo inferred by calibrating a 30 cm PNeo image using Sentinel-2 imagery using the calibration transfer method.
\end{enumerate}

We began by evaluating the two methods of albedo inference: convolution calibration and calibration transfer. Convolution calibration uses a circular averaging kernel of radius 10 meters (the size of a Sentinel-2 pixel) to rescale each band of the high-resolution image by the ratio of the kernel convolutions of the calibrated and uncalibrated image bands. Albedo is then computed from the calibrated high-resolution bands. The alternative method, calibration transfer, uses a machine learning model trained (on an image-wise basis) on upscaled high-resolution imagery and targeting Sentinel-2 albedo to infer the high-resolution albedo.

We evaluated these methods beginning with the most precise data: surface albedo estimates for roof segments generated from a single Sentinel-2 image (April 30, 2024) and a single PNeo image (May 8, 2024), the high-resolution image being acquired on the same date as the NEON flyover and the Sentinel-2 image being the cloud-free image most proximate. This limited, high-precision ground truth sample contains 440 roof segments. Compared to NEON ground truth data, the calibration transfer method produces a slightly higher R-squared value than the convolution-based calibration, although its RMSE and bias are also marginally greater (Table~\ref{tab:methods-compare}). This pattern holds whether the analysis is conducted on uncorrected (image albedo) or slope-corrected (surface albedo) reflectance values derived from individual roof segments. Despite these corrections, the accuracy of the surface albedo and the image albedo are nearly equivalent when validated against NEON reference data, with the goodness-of-fit metrics of the image albedo exceeding those of the surface albedo for May 23, 2023.

\begin{table}[t]
\centering
\caption{Goodness of fit metrics comparing the convolution calibration, calibration transfer, and the Sentinel-2 albedo estimates to the May 8, 2024 NEON ground truth data. The albedos are calibrated from a single, cloud-free Sentinel-2 image from April 30, 2024.
}
\label{tab:methods-compare}
\resizebox{\textwidth}{!}{%
\begin{tabular}{lclrrrrrrrrr}
\toprule
 &  &  & \multicolumn{3}{c}{$R^2$} & \multicolumn{3}{c}{RMSE} & \multicolumn{3}{c}{Bias} \\
\cmidrule(lr){4-6} \cmidrule(lr){7-9} \cmidrule(lr){10-12}
High-res Date & N & Albedo Type & Convolution & Transfer & Sentinel-2 & Convolution & Transfer & Sentinel-2 & Convolution & Transfer & Sentinel-2 \\
\midrule\addlinespace[2.5pt]
May 5, 2023 & 11202 & Image   & 0.818 & 0.837 & 0.678 & 0.049 & 0.056 & 0.065 & 0.007 & 0.032 & 0.017 \\
May 5, 2023 & 11202 & Surface & 0.813 & 0.828 & 0.654 & 0.050 & 0.059 & 0.067 & 0.010 & 0.035 & 0.020 \\
May 8, 2024 & 440   & Image   & 0.918 & 0.927 & 0.767 & 0.046 & 0.048 & 0.080 & -0.004 & 0.018 & -0.007 \\
May 8, 2024 & 440   & Surface & 0.920 & 0.927 & 0.762 & 0.046 & 0.049 & 0.081 & -0.003 & 0.019 & -0.006 \\
May 16, 2024 & 15396 & Image   & 0.891 & 0.900 & 0.674 & 0.034 & 0.038 & 0.063 & 0.006 & 0.015 & 0.019 \\
May 16, 2024 & 15396 & Surface & 0.892 & 0.898 & 0.637 & 0.034 & 0.039 & 0.066 & 0.009 & 0.018 & 0.023 \\
\bottomrule
\end{tabular}
}
\end{table}

We expand our sample by also comparing the results of the two methods using the high-resolution images from May 23, 2023, and May 16, 2024. As the temporal gap between the high-resolution imagery and the NEON flyover increase, the goodness-of-fit of both methods decreases, although the relative performance of the methods remains the same, with the calibration transfer performing slightly better than the convolution calibration, and both inference methods outperforming the Sentinel-2 albedo estimates. Both the convolution calibration and calibration transfer methods are effective at filling in the details of the inferred albedo image by making use of the high resolution content of the uncalibrated image. These are only two possible models, notably there are in theory an infinite number of possible high-resolution albedo images that will radiometrically match Sentinel-2 when they are upscaled to 10 meter resolution. It is not a goal of this study to exhaustively explore the space of all calibration methods but rather to pick two rationally constructed yet algorithmically very different models and compare these to ground truth comprising the aerial NEON albedo. As the results show, two methods, an image processing approach and a machine learning approach, give very similar results even though the methods differ significantly. However, the calibration transfer method is a more complicated and computationally expensive method of radiometrically calibrating the high-resolution imagery, and the marginal improvement in accuracy is not a meaningful gain over the simple method of convolution calibration. Based on this validation and the relative ease of computation we choose convolution calibration as the preferred method of albedo inference and we use this method for the illustrative application that follows.

The availability of cloud-free Sentinel-2 imagery proximate in time to high-resolution imagery is not guaranteed; additionally, changing solar elevation and ground conditions impact Sentinel-2 albedo estimates. To minimize these concerns, we calculate the temporal median albedos from the 50 days on either side (100-day window) of the high-resolution images to use as the calibration source for the high-resolution albedo inferences. We then compare the performance of the albedo inferences against the Sentinel-2 albedo estimates when validated against the NEON ground truth (Table~\ref{tab:gof}). The temporal medians are slightly less accurate compared to NEON, but still perform strongly. Like with the single-day inferences, for both the high-resolution albedo and Sentinel-2 imagery as the temporal gap between the NEON flyover and the centers of the temporal medians increases, the accuracy decreases. This is true regardless of whether or not geometric corrections have been made. Across all metrics and temporal resolutions, the high-resolution albedo outperforms Sentinel-2.  Intercomparison results are similar regardless of whether one is comparing image albedo or surface albedo.

\begin{table}[t]
\centering
\caption{Goodness of fit metrics comparing the single day and temporal median albedos from both the high-resolution albedo inferences and Sentinel-2 albedo estimates to the May 8, 2024 NEON ground truth data. The temporal medians are calculated from the 50 days on either side of the date of the high-resolution imagery.}
\label{tab:gof}
\resizebox{\textwidth}{!}{%
\begin{tabular}{lrlrrrrrrrrrrrr}
\toprule
 &  &  & \multicolumn{2}{c}{Downscaled} & \multicolumn{2}{c}{Sentinel-2} & \multicolumn{2}{c}{Downscaled} & \multicolumn{2}{c}{Sentinel-2} & \multicolumn{2}{c}{Downscaled} & \multicolumn{2}{c}{Sentinel-2} \\
 \cmidrule(lr){4-5} \cmidrule(lr){6-7} \cmidrule(lr){8-9}
\cmidrule(lr){10-11} \cmidrule(lr){12-13} \cmidrule(lr){14-15}
 &  &  & \multicolumn{4}{c}{$R^2$} & \multicolumn{4}{c}{RMSE} & \multicolumn{4}{c}{Bias} \\
\cmidrule(lr){4-7} \cmidrule(lr){8-11} \cmidrule(lr){12-15}
High-res date & n & Albedo & 
\shortstack{Single\\date} & \shortstack{Temporal\\median} & 
\shortstack{Single\\date} & \shortstack{Temporal\\median} & 
\shortstack{Single\\date} & \shortstack{Temporal\\median} & 
\shortstack{Single\\date} & \shortstack{Temporal\\median} & 
\shortstack{Single\\date} & \shortstack{Temporal\\median} & 
\shortstack{Single\\date} & \shortstack{Temporal\\median} \\
\midrule\addlinespace[2.5pt]
May 5, 2023 & 11,202 & Image   & 0.818 & 0.808 & 0.678 & 0.644 & 0.049 & 0.059 & 0.065 & 0.079 & 0.007 & 0.026 & 0.017 & 0.037 \\
May 5, 2023 & 11,202 & Surface & 0.813 & 0.795 & 0.654 & 0.598 & 0.050 & 0.062 & 0.067 & 0.083 & 0.010 & 0.029 & 0.020 & 0.041 \\
May 8, 2024 & 440    & Image   & 0.918 & 0.910 & 0.767 & 0.762 & 0.046 & 0.051 & 0.080 & 0.085 & -0.004 & 0.004 & -0.007 & 0.003 \\
May 8, 2024 & 440    & Surface & 0.920 & 0.911 & 0.762 & 0.754 & 0.046 & 0.051 & 0.081 & 0.087 & -0.003 & 0.005 & -0.006 & 0.004 \\
May 16, 2024 & 15,396 & Image   & 0.891 & 0.894 & 0.674 & 0.676 & 0.034 & 0.042 & 0.063 & 0.075 & 0.006 & 0.024 & 0.019 & 0.040 \\
May 16, 2024 & 15,396 & Surface & 0.892 & 0.888 & 0.637 & 0.606 & 0.034 & 0.045 & 0.066 & 0.080 & 0.009 & 0.028 & 0.023 & 0.045 \\
\bottomrule
\end{tabular}
}
\end{table}

\begin{figure}
    \centering
    \includegraphics[width=1\linewidth]{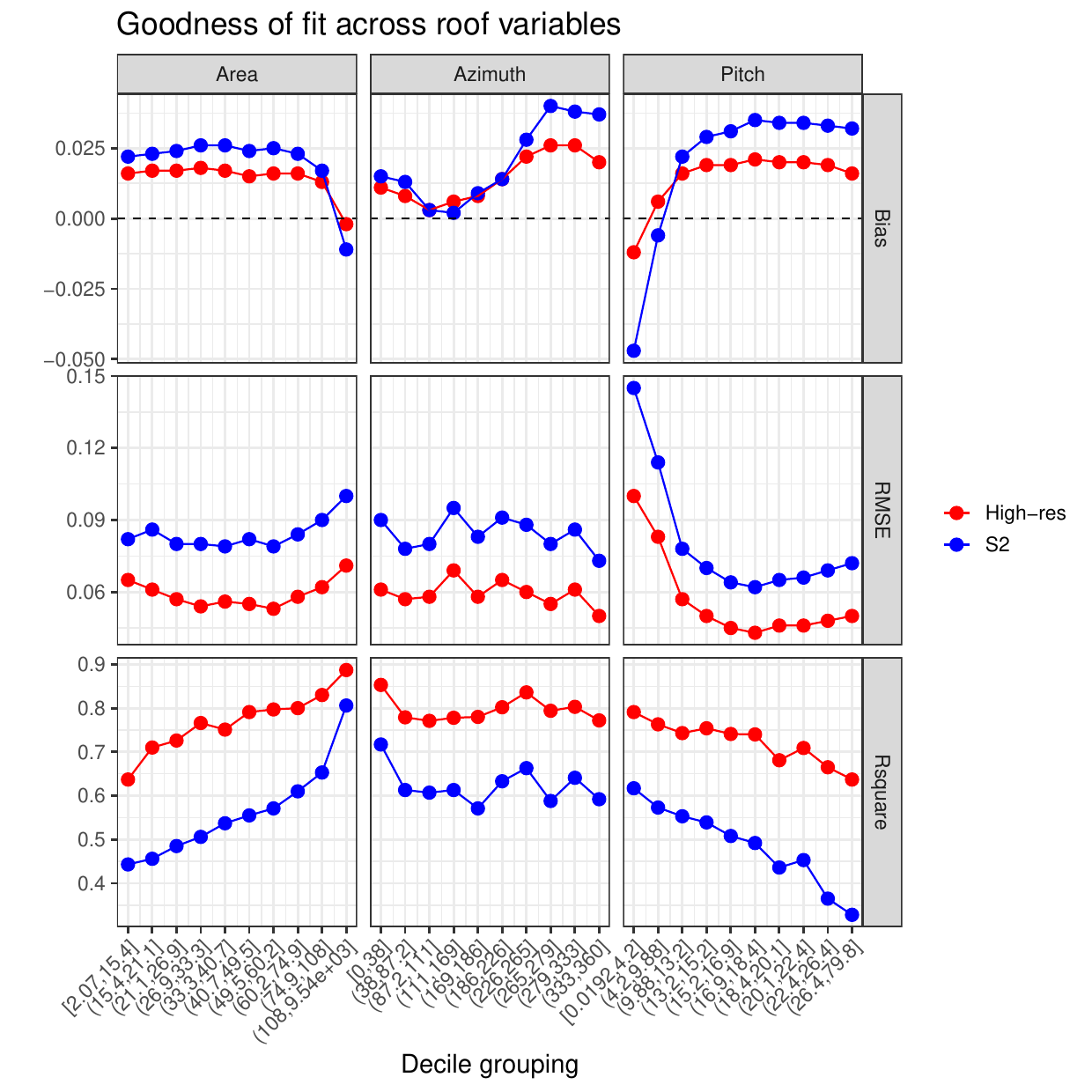}
    \caption{Roof level variables are binned into deciles and the goodness-of-fit calculated for each dataset (high-resolution albedo inference in red and Sentinel-2 albedo in blue) against the NEON ground truth data. From left to right, the columns show the goodness-of-fit for roof area ($m^2$), azimuth, and pitch (slope). From top to bottom, the rows show the bias, RMSE, and $R^2$ metrics. The high-resolution albedo is more robust to the effects of variation in roof geometry and size.
}
    \label{fig:gof}
\end{figure}

To assess how surface and solar geometry impacts the accuracy of the data we examine the goodness of fit of the uncorrected image albedos by roof segment pitch, roof segment azimuth, and roof segment area  (Fig.~\ref{fig:gof}). We bin the pitch, azimuth, and area of the roof segments into deciles and calculate the $R^2$, RMSE, and bias values for the observations in each bin. For each roof segment variable, the convolutional high-resolution albedo outperforms Sentinel-2, with a bias closer to zero, lower RMSE values, and higher $R^2$ values. In particular, the high-resolution albedo inferences provide better inferences for small roof segments and are more robust to the effects of increased pitch. Expectedly, as the roof segment area increases, the Sentinel-2 imagery approaches the accuracy of the high-resolution inferences.

\begin{figure}
    \centering
    \includegraphics[width=1\linewidth]{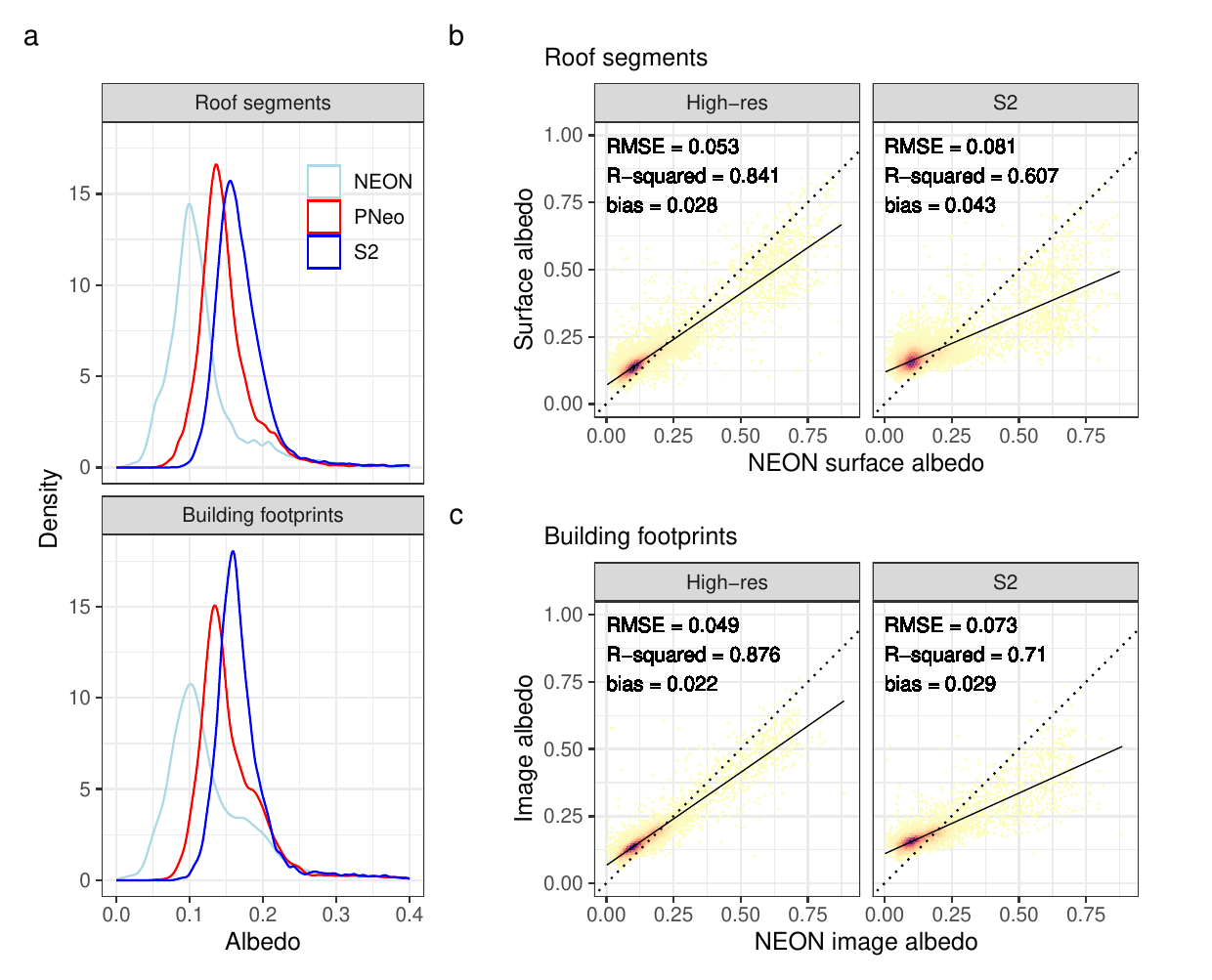}
    \caption{Geometrically corrected surface albedo of roof segments (top a, b) and the uncorrected image albedo of the building footprints (bottom a, c) for each dataset compared to the NEON ground truth data. The high-resolution albedo inferred from convolutional calibration performs better than Sentinel-2 by all measures and more closely matches the distribution of the NEON albedo.
}
    \label{fig:dens}
\end{figure}

Finally, we examine the overall accuracy of the two datasets, including all roof segments from all images (n = 54,076)  (Fig.~\ref{fig:dens}). For the roof segments, both the high-resolution albedo inference and the Sentinel-2 albedo are biased high relative to the NEON albedo, in general overestimating albedos. However, the density plots show that the convolution calibration methods (top PNeo distribution in Fig.~\ref{fig:dens}) effectively shifts the distribution of the Sentinel-2 albedo towards that of the NEON ground truth. The albedo inference explains almost 25\% more of the variation in the NEON albedo than Sentinel-2 ($R^2$ value of 0.841 compared to 0.607), has a lower RMSE, and is less biased.

While the roof segment outlines provide valuable information on pitch and azimuth, they are not yet publicly available. In contrast, the building footprints provided by Overture Maps are globally available and aggregating albedo to the roof level is a likely use case for the data  (Fig.~\ref{fig:maps}). We evaluate the accuracy of the convolutional  albedo inference and the Sentinel-2 albedo against the NEON albedo measurements when aggregated to the building footprint (Fig.~\ref{fig:dens}). We calculate the median albedo for each building footprint in the study area (n = 10,767) for each dataset. Where a building footprint is covered by multiple NEON flightlines, we use the albedo value from the flightline with the higher solar elevation to more closely match the solar conditions of the satellite overpasses (63.6° solar elevation for the PNeo acquisition and 61.46° solar elevation for the Sentinel-2 overpass on May 8, 2024). The density plot (lower, Fig. 4) shows that in addition to overestimating building footprint albedos, Sentinel-2 also regresses towards the mean, with a narrower distribution than observed in NEON. The convolution calibration methods both shift the distribution to the left and widen the distribution, making it more similar to that of the NEON albedo measurements. The goodness of fit is higher than for the roof segments, unsurprising because the values are averaged over larger areas than for the roof segments. While the accuracy of Sentinel-2 is higher for building footprints than for the roof segments ($R^2$ of 0.88 compared to 0.71), the high-resolution albedo inference still provides a considerable increase in accuracy, explaining almost 17\% more of the building level variation in NEON albedo.

\subsection{Boulder cool roof scenario evaluation}

Having validated our methods for estimating the albedo of building footprints, we apply these methods to quantify the solar radiation management potential for Boulder. Specifically, we calculate the potential albedo change (albedo delta) for the building footprints if they were converted to cool roofs using a target albedo value of 0.55. This target value represents a conservative aged albedo achievable using existing and widely available cool roofing materials and was chosen based on the ENERGY STAR cool roof albedo. For buildings with existing albedos exceeding the target value, their albedo potential is defined as zero because the building is already considered to already have a cool roof. To quantify the potential contribution of a cool roof to the overall area albedo, we multiply the albedo potential by the area of the roof, which we term the effective area. This provides a measure of solar radiation management potential in units of area corresponding to the 100\% reflective material equivalent. By summing the effective areas and then normalizing by the area of the city boundary, we can estimate the increase in city-wide average albedo that would be achieved by adoption of cool roofs. Using this approach, we calculated the effective area of each building footprint then estimated the overall increase in city-wide albedo. We additionally estimated the reduction in air temperature associated with increased albedo by using the average albedo cooling effectiveness (ACE) of 0.4°C per 0.10 albedo increase, the median value from the \cite{krayenhoff2021} synthesis.

We evaluated three scenarios, using the intersection of the Boulder city limits and the building footprints covered by the NEON flightlines as the city boundary:

\begin{enumerate}[label=\Roman*.]
\item All roofs converted to cool roofs
\item Large buildings (building footprints above the 90th percentile of area) converted to cool roofs 
\item Buildings with the lowest albedos (buildings above the 90th percentile of albedo delta) converted to cool roofs 
\end{enumerate}

Scenario I estimates the maximum albedo change possible for Boulder through cool roof implementation, and while likely not realistic, gives a sense of what is technically possible. Scenario II prioritizes buildings with large footprint areas. These buildings are very likely to have flat slope roofs making cool roofs a cost effective option and the data shows that implementation will have a large impact (SI Figure 1). Scenario III prioritizes the buildings with the greatest albedo potentials, meaning they have the lowest current albedos. Implementing cool roofs on low albedo buildings would reduce energy costs and interior temperatures, and a policy that prioritizes the darkest buildings could improve heat equity. 

\begin{figure}
    \centering
    \includegraphics[width=0.5\linewidth]{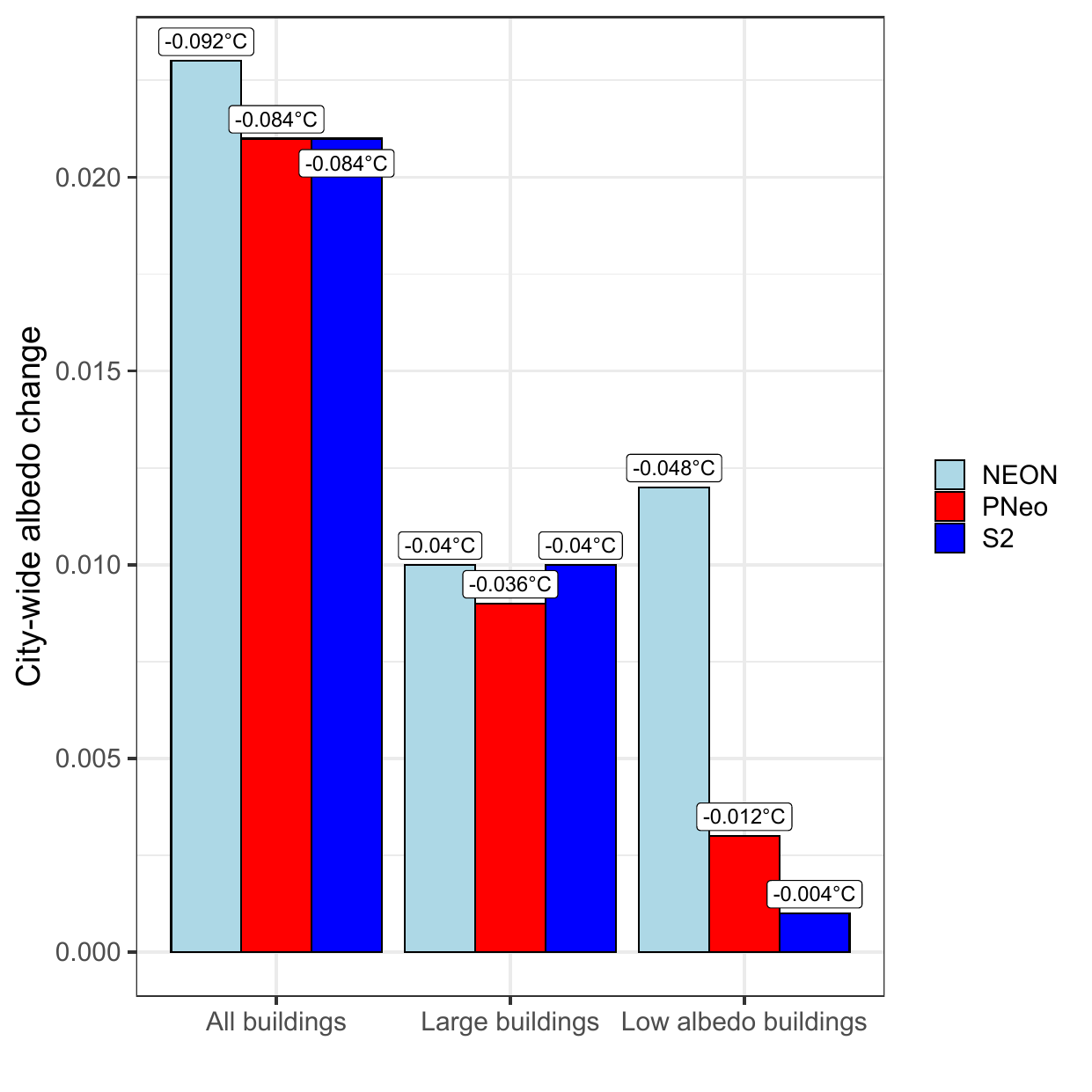}
    \caption{The bars represent the albedo change estimated by each dataset for each of the three scenarios in Boulder, CO. Labels indicate the temperature reduction associated with the albedo change. }
    \label{fig:boulder-scenarios}
\end{figure}

\begin{table}[t]
\centering
\caption{Implementation scenarios for Boulder, CO.}
\label{tab:boulder-scenarios}
\resizebox{\textwidth}{!}{%
\begin{tabular*}{\linewidth}{@{\extracolsep{\fill}}clrrrrr}
\toprule
Scenario & Sensor & N & \makecell{Percent of\\city area} & \makecell{Effective\\area ($m^2$)} & \makecell{City-wide\\albedo $\Delta$} & \makecell{Temperature\\$\Delta$ (°C)} \\
\midrule\addlinespace[2.5pt]
I        & NEON & 10378 & 0.066 & 1104803.45 & 0.023 & -0.092 \\
I        & PNeo & 10550 & 0.066 & 1044323.32 & 0.021 & -0.084 \\
I        & S2   & 10607 & 0.066 & 1039107.80 & 0.021 & -0.084 \\
II      & NEON & 1063  & 0.036 & 478880.26  & 0.010 & -0.040 \\
II      & PNeo & 1063  & 0.036 & 456381.32  & 0.009 & -0.036 \\
II      & S2   & 1063  & 0.036 & 466775.93  & 0.010 & -0.040 \\
III & NEON & 5232  & 0.025 & 577433.40  & 0.012 & -0.048 \\
III & PNeo & 1064  & 0.006 & 126771.52  & 0.003 & -0.012 \\
III & S2   & 140   & 0.002 &  44991.73  & 0.001 & -0.004 \\
\bottomrule
\end{tabular*}
}
\end{table}

While most urban areas have roof fractions around 20\%, the NEON data was collected over a largely suburban area. To account for this, the boundary used to estimate the impact of the cool roof scenarios is the intersection of the city boundary and the overture buildings for which we have NEON albedos. The area of this boundary is 48,848,660 $m^2$ and contains 10,635 buildings with a total footprint area of 3,235,055 $m^2$, or approximately 6.6\% of the land area. Estimates for city-wide albedo change are low because Boulder is not densely urbanized and the data only covers part of the urban area.

The greatest change in albedo and therefore the largest temperature reductions come from Scenario I (Table~\ref{tab:boulder-scenarios}). The effective areas calculated from the three datasets vary because they estimate different building albedos, however, when normalized over the city area the results are not meaningfully different. The large buildings scenario implements cool roofs on buildings with areas greater than the 90th percentile (398 $m^2$). Like Scenario I, the differences in effective area are insignificant when normalized by the city area. However, for Scenario III, which implements cool roofs based on low existing albedos (albedo potential greater than the 90th percentile value of 0.435, meaning that cool roof implementation would increase the albedo of a building by at least this amount), the differences are much greater. While NEON finds 6855 high potential buildings, the high-resolution inference identifies 20\% of this number, and Sentinel-2 only identifies less than 3\% of this number. Although the convolution calibration method reduces the overestimation of low albedos, the results do not fully correct for the positive bias of Sentinel-2. We hypothesize that this is primarily due to undercorrection of the Sentinel-2 surface reflectance for atmospheric scattering, as highlighted in the Discussion section.

\subsection{Global cities cool roof scenarios}

\begin{figure}
    \centering
    \includegraphics[width=1\linewidth]{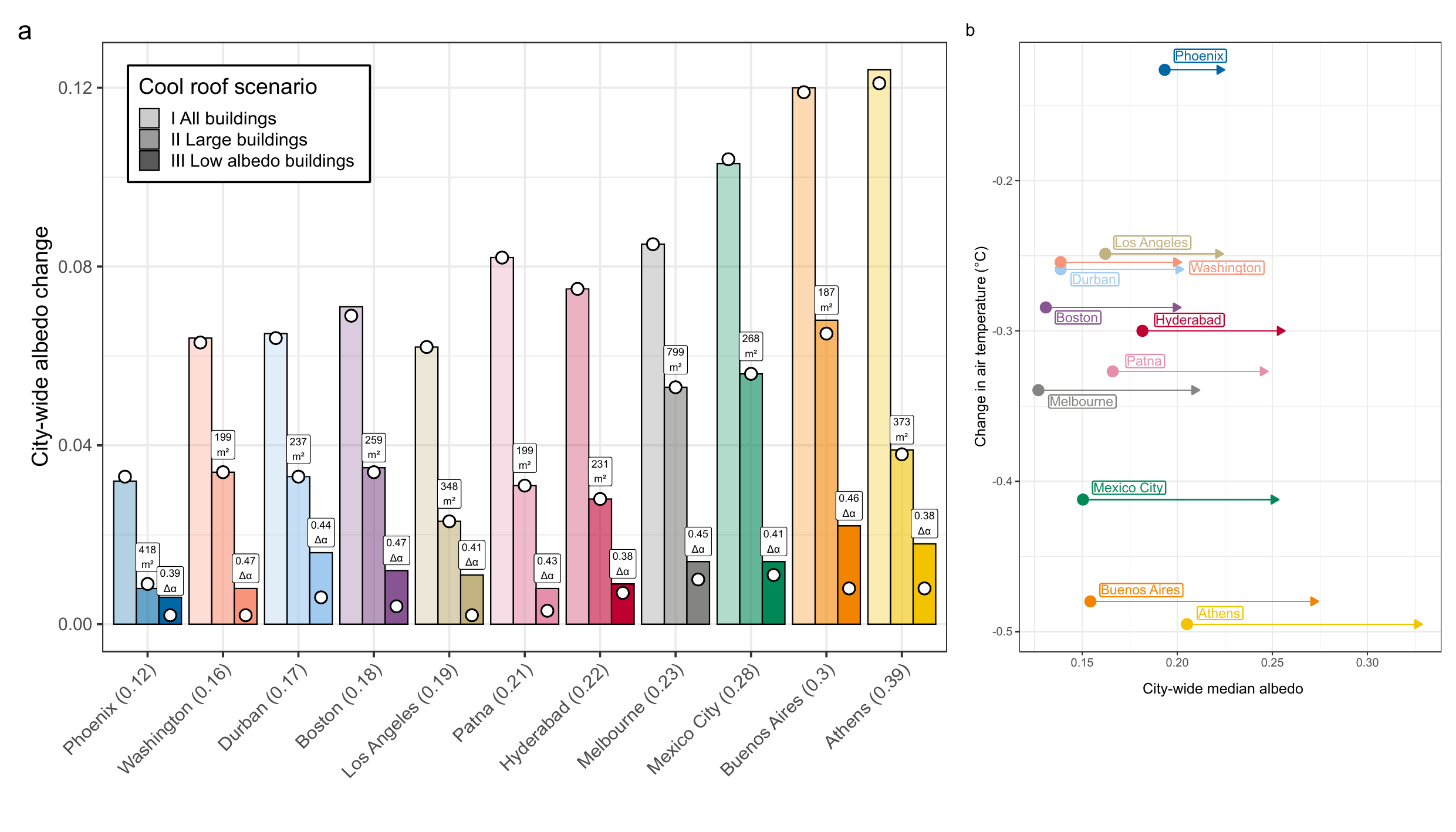}
    \caption{(a) Bars represent estimates for the city-wide albedo change from three cool roof scenarios calculated from the high-resolution albedo inferences. The circles show the estimates derived from the Sentinel-2 data. The numerical values above the bars are the threshold values for prioritization for each scenario derived on a per city basis. The cities are ordered by the percent of the city area covered by buildings, with the lowest on the left. This percent coverage is also indicated on the x-axis labels. (b) The temperature reduction associated with city-wide cool roof implementation (Scenario I) derived using the albedo cooling efficiency from \cite{krayenhoff2021}. The current city-wide median albedo is represented with a circle and the albedo resulting from cool roof adoption is shown with the arrow. 
}
    \label{fig:global-scenarios}
\end{figure}

We extend these methods for estimating the solar radiation management potential to a selection of 12 global cities (Fig.~\ref{fig:global-scenarios}). These cities were chosen to represent a range of climatic, geographic, morphological, and economic conditions. We acquired the Overture Maps building footprints that intersect the city geometries as well as the available PNeo imagery for the cities. We generate both the Sentinel-2 albedo estimates and the high-resolution albedo inference using the 100-day medians centered on the available PNeo dates and then calculate the median albedo of each building footprint for each image. Where there are multiple images (either Sentinel-2 or PNeo) for a building footprint we take the median value so that we have one estimate each for the two albedo datasets per building footprint. Additionally, we calculate the area of each building footprint. We follow the same methods for calculating the cool roof scenarios as for Boulder.

The albedo change associated with Scenario I is the greatest for every city and is strongly dependent on the building fraction (Fig.~\ref{fig:global-scenarios}). However, in many cases Scenario II accounts for over half of the increase from Scenario I, showing that prioritizing large buildings could be a highly effective policy for a city to pursue. The results from Scenario III show that prioritizing buildings with high albedo deltas, while potentially having positive effects for equity, has a much lower impact on the city-wide albedo. However, it is important to note that our experiment in Boulder shows that both the Sentinel-2 and the high-resolution inference fail to identify many buildings in this scenario, with the ground truth albedo identifying five times as many buildings with high albedo deltas and estimating a higher albedo change from Scenario III than Scenario II. It is possible that the results for other cities follow a similar pattern. The high-resolution inferences and the Sentinel-2 estimates predict quite similar albedo increases for Scenarios I and II, however, they predict very different increases for Scenario III. We hypothesize that this is due to the inability of Sentinel-2 to accurately resolve the albedo of smaller buildings due to its coarse spatial resolution. Importantly, the greater the increase in city-wide albedo, the greater the potential temperature reduction—our results show that cities could achieve up to a 0.5°C cooling effect with full-scale implementation of cool roofs.

\section{Discussion}

Our convolution calibration method combines the spatial information of high resolution imagery with the spectral and radiometric information from Sentinel-2 to create highly accurate estimates of rooftop albedo at a 30 cm resolution. Validating albedo data is a considerable challenge and our novel use of the NEON data allows us to evaluate the ability of our albedo inferences to accurately estimate the albedo of urban roofs. Sentinel-2 albedo is the highest resolution albedo data currently available to the public, and we show that while it adequately measures the spatial variation in urban albedo at city scales and for large buildings, it overestimates the albedo of dark roofs and underestimates the albedo of light roofs due to the scale mismatch between urban features and the 10 meter spatial resolution. The high-resolution inference methods we present here reduce this tendency by broadening the distribution of albedo values (Fig. 4). By all measures, the high-resolution inferences do a better job of estimating the ground truth albedo of roofs as measured by the NEON hyperspectral instrument than Sentinel-2.

Both Figures 1 and 4 illustrate that, in comparison to NEON, the inferred albedo underestimates high albedos and overestimates the low albedos. One possible explanation for this stems from the original Sentinel-2 surface reflectance data which is used for calibration. The Sentinel-2 surface reflectance product relies on atmospheric correction which attempts to remove the effects of the atmosphere (gases, aerosols, water vapor) to provide a more accurate representation of the surface reflectance. Two main effects from the atmosphere are: (1) scattering of extraneous light from the bulk of the atmosphere into the imager; and (2) scattering within the direct line of sight through the atmosphere column that prevents light leaving the surface from reaching the imager. These effects are corrected to first order by subtracting an offset from the albedo and applying a positive gain to the albedo, respectively. New versions of SEN2COR are regularly released and deployed as new processing baselines; as methods improve, calibration accuracy is expected to improve commensurately.

This experiment suggests that for this particular set of conditions during the NEON flyover, the agreement with NEON would be improved by a gain and offset correction to the Sentinel-2 image, suggesting that SEN2COR may have applied too small of a correction of both extraneous atmospheric light scattering and atmospheric column scattering. While we could have fit such a correction to this experiment, we elected to show the results as is because they represent how the inference is applied in general where we do not have ground truth data. With this single experiment, it is not possible to know whether undercorrection for the scattering effects described occurs systematically or stochastically, but hyperspectral measurements like those from NEON offer a powerful tool for evaluation.

Using the high-resolution albedo inferences to estimate the median albedo of roofs in 12 global cities, we estimate that cities can see up to a 0.5°C cooling effect from full scale implementation of cool roofs. We also illustrate the utility of robust albedo estimates associated with building footprints to estimate the effects of different cool roof policy scenarios. We show that prioritizing the largest buildings is a highly effective policy pathway for reducing heat by implementing cool roofs. While the high-resolution albedo and the Sentinel-2 albedo produce similar results when estimating the albedo increase from cool roof implementation on all buildings (Scenario I) and large buildings (Scenario II), they produce quite different results at finer spatial scales, like when prioritizing buildings based on their existing albedo (Scenario III). The 10-m resolution of Sentinel-2 is inadequate to measure the tails of the existing building albedo distribution---consequently it fails to identify many buildings that would see the most improvement with a cool roof (high albedo potential). While the high-resolution albedo may still miss the extremes of the tails, it provides a stronger capability for estimating the albedo of very dark or very light buildings. Critically, the city-wide change in albedo varies across the cities based on both their current albedos and the distribution of building sizes. For instance, in some cities the largest buildings are also the brightest, yielding a lower delta. 

\begin{figure}
    \centering
    \includegraphics[width=1\linewidth]{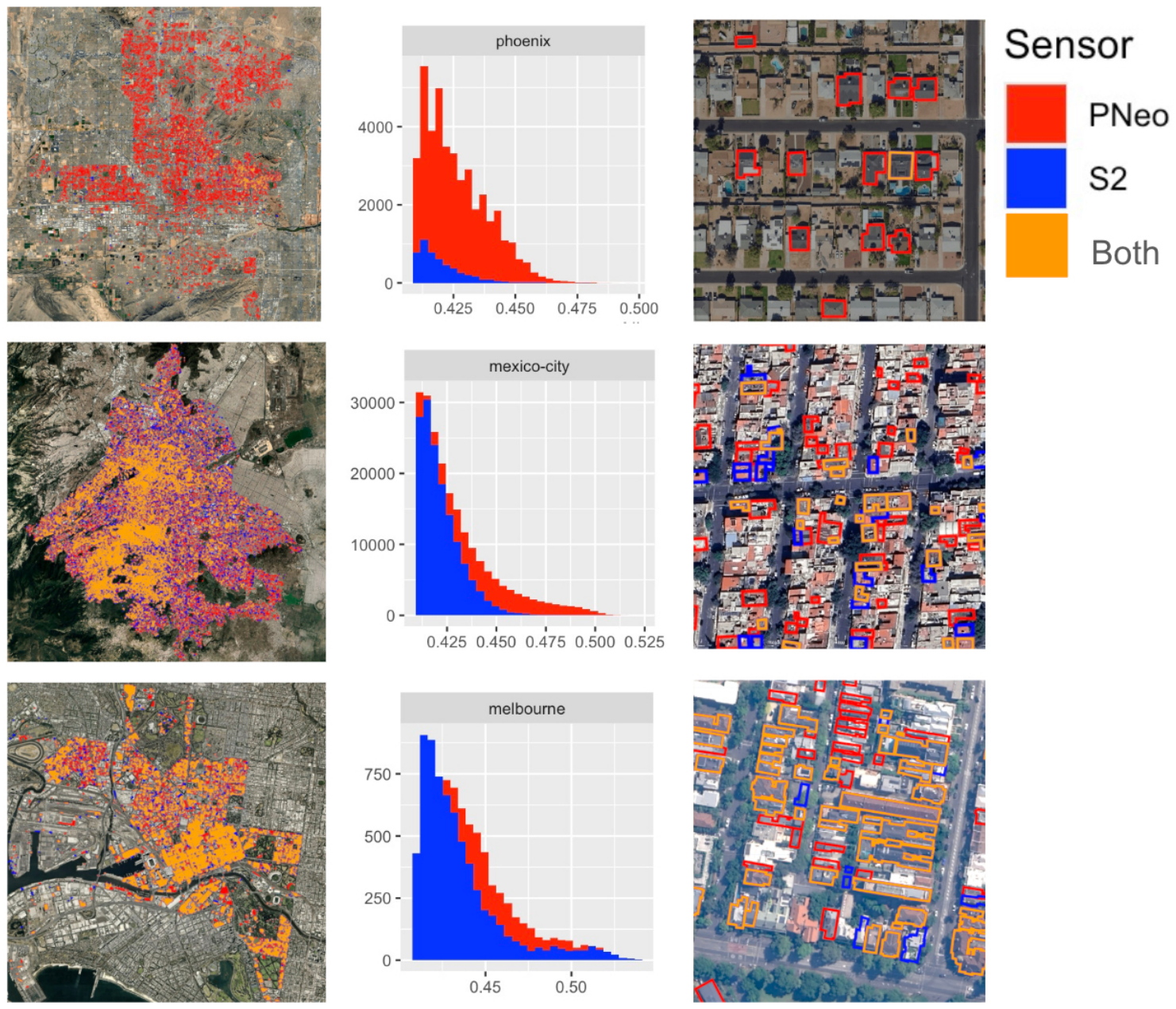}
    \caption{Buildings identified as having an albedo potential of above 0.41 in the high-resolution albedo in blue, Sentinel-2 in red, and in both in orange. This threshold is the albedo potential value at which the percent of the buildings identified from the high-resolution albedo that are also identified from the Sentinel-2 albedo dropped below 50. The albedo potential is the amount the albedo of a roof would increase with cool roof implementation. The panel on the left shows the city-wide spatial distribution of the selected buildings, the middle panel shows histograms of the building albedo potentials higher than the threshold, with the number of buildings in each bin represented by the y-axis, and the right panel shows zoomed in areas of each city. Some cities have high agreement between the datasets as in the case of Melbourne, while other cities show poor agreement as in the case of Phoenix. 
}
    \label{fig:final}
\end{figure}

Both the data validation and the global cities analysis highlight that there are multiple use cases for urban albedo data, one of which can be adequately filled by the widely available Sentinel-2 data and another which requires higher resolution data like we present in this paper. When estimating the albedo increase of implementation from scenarios that do not involve prioritizing buildings based on their existing albedo—cool roofs on all buildings or cool roofs on large buildings—the high-resolution albedo and Sentinel-2 give very similar estimates (Fig.~\ref{fig:global-scenarios}a). Indeed, when aggregating across many buildings the high-resolution albedo and the Sentinel-2 albedo yield similar estimates because errors cancel and the area of buildings as a fraction of the urban area is the same for both (SI Fig. 2). We conclude that for a general research use case, the Sentinel-2 data does a good job of estimating the albedo change from cool roof implementation at the neighborhood scale and larger. However, for a use case that involves the prioritization of individual buildings, the high-resolution albedo and the Sentinel-2 albedo tell very different stories (Fig.~\ref{fig:final}). The relatively coarse spatial resolution of Sentinel-2 causes a regression to the mean, underestimating the albedo of existing high albedo surfaces, and overestimating the albedo of dark surfaces. For urban planning use cases that require understanding variation at the individual building scale, then, the high-resolution albedo is necessary to prioritize buildings based on their current albedo. 

\subsection{Limitations}

This study was validated exclusively in Boulder, Colorado, which may limit the generalizability of the results. Boulder’s built environment includes unique characteristics such as specific roof geometries, materials, and building height restrictions that may not be representative of many other urban areas. In particular, the city’s stringent height ordinances result in a lack of the kinds of high-rise structures common in denser metropolitan contexts. Additionally, the building density of our validation area is much lower (approximately 6\%) than typical urban areas (approximately 20\%). The analysis was further constrained by the spatial extent of the airborne data—NEON focuses primarily on agricultural regions and so did not cover all of Boulder. This reduced the size and diversity of the validation dataset. Additionally, the surface properties of Boulder’s natural environment, including coloration and bidirectional reflectance (BRDF), are likely to differ from those in other geographic regions. Atmospheric conditions in Boulder may also affect transferability. The city’s high elevation and generally low aerosol concentrations yield an atmospheric profile with radiative properties distinct from lower-altitude or more polluted locations, potentially influencing the observed reflectance characteristics.

\section{Conclusions}

Our findings demonstrate that Sentinel-2, despite its 10-meter resolution, provides an effective and globally accessible means of estimating urban albedo. Somewhat unexpectedly, we also show that, in aggregate, it can offer useful estimates of citywide albedo impacts from cool roofs. However, decisions about cool roof interventions are typically made building-by-building; here, the high resolution albedo inferences provide the critical actionable information.

We find that high-resolution RGBN imagery, even if uncalibrated, can be leveraged to infer rooftop albedo with an error of approximately 4\% through calibration against contemporaneous Sentinel-2 imagery. The frequent revisit time of Sentinel-2 makes this calibration feasible in most locations, although persistent cloud or snow cover may pose challenges. When accurate digital surface models are available, further correction for roof pitch and solar angles can produce albedo estimates that more accurately represent the intrinsic optical properties of rooftops rather than snapshot imaging conditions. Among sources of high-resolution imagery, Pléiades Neo offers a key advantage by sharing a similar sun-synchronous orbit with Sentinel-2, minimizing illumination differences between images captured under near-nadir conditions. For other high-resolution imagery sources, additional steps to harmonize the differing shadows in the two images may be necessary to achieve similar precision.

Finally, while our validation against NEON airborne data demonstrated strong agreement with satellite-based estimates, inherent differences in acquisition timing and shadow geometry underscore the challenges of exact cross-sensor comparisons. Additionally, due to the slow speed of low-flying aircraft, aerial imagery cannot be acquired contemporaneously with broad area satellite imagery. Our May 8, 2024 experiment suggests that residual scattering effects may contribute to undercorrection in Sentinel-2 reflectance retrievals, a consideration that should be further explored in future studies.

Together, these results provide a practical and scalable approach for high-precision urban albedo mapping and emphasize the critical role of high-resolution calibrated imagery for targeted climate interventions such as cool roof deployment.

\section{Methods}

\subsection*{Calibration imagery}

We use the atmospherically corrected Sentinel-2 Level 2A (Sentinel-2) surface reflectance dataset as the source of calibration \cite{ESA}. Sentinel-2 was chosen because of its relatively high spatial resolution of 10 meters, compared to other radiometrically calibrated satellite data products like LANDSAT (30 m) and MODIS (1 km). Sentinel-2 also has a regular revisit time (3-5 days) and consistent operation close to nadir, with redundancy achieved with its 2-satellite constellation. The Sentinel-2 dataset is publicly and globally available and is maintained as an asset in Earth Engine \url{(https://developers.google.com/earth-engine/datasets/catalog/COPERNICUS_Sentinel-2_SR_HARMONIZED)}. Because we are inferring solar spectrum albedo, we model the Sentinel-2 bands which cover the solar spectrum: B2, B3, B4, B8, B11 and B12. As was originally done by \cite{bonafoni2020} we checked the broad spectrum solar albedo computation against the network of SURFRAD stations located in the United States. 

\subsection*{Artifact masking and mosaicking}
Using every Sentinel-2 image tile that overlaps the high-resolution input image over a 100-day window centered on the date of the high-resolution image acquisition, we create a temporal median mosaic from Sentinel-2 imagery masked using the Cloud Score Plus dataset \cite{pasquarella2023} and the Sentinel-2 snow probability mask. Since clouds are unlikely to repeatedly appear in the same place every time a satellite flies overhead, a sufficiently large number of images will ensure full coverage of the high-resolution input image, and the median statistic is effective at eliminating any unmasked cloud and cloud-shadow artifacts. 

\subsection*{High resolution input imagery}
The geospatial input data used to infer solar albedo is Pléiades Neo (Pléiades Neo 3 and Pléiades Neo 4: © 2024 CNES / Airbus) satellite  imagery. The resolution of the input data (30 cm) is substantially higher than the resolution of the Sentinel-2 calibration data (10 m). The input imagery was taken by tasked satellite flyovers. This input imagery is a top of atmosphere measurement, and it does not typically have as much solar spectrum coverage as the calibration data. Before use as input data, the input imagery is subjected to processing including orthorectification and pansharpening.

\subsection*{Coregistration}
The European Space Agency reports Sentinel-2 pointing accuracy to be about 20 meters \url{https://www.esa.int/Enabling_Support/Operations/Sentinel-2_operations} and our own visual inspection confirmed that each Sentinel-2 flyover image will have pixels that jitter with respect to other flyover images by about ±10 meters. To avoid loss of resolution when we make composite images from multiple flyovers, we use our high resolution input imagery as a reference to which we co-register \url{https://developers.google.com/earth-engine/guides/register} the Sentinel-2 image mosaic before conducting additional processing. 

\subsection*{Convolution calibration}
A mathematical method to increase the spatial resolution (downscale) of the calibrated Sentinel-2 image is to apply a circular kernel convolutional calibration. The high resolution B, G, R, N, N, and N bands are calibrated using the B2, B3, B4, B8, B11 and B12 bands of Sentinel-2 respectively. Because there is only one infrared band in the high resolution image, it is replicated threefold. In this method, a circular averaging kernel, $k_c(x,y)$ of radius 10 meters (the size of a Sentinel-2 pixel) is convolved, band-for-band with both the high resolution image $h(x,y)$ and the low resolution image $l(x,y)$. The calibrated image $c(x,y)$ is then computed from:

\begin{equation}\label{conv_eqn}
    c(x,y) = k_c(x,y)\otimes l(x,y) \cdot h(x,y) / kc(x,y) \otimes h(x,y) 
\end{equation}

This quotient of convolutions guarantees that $k_c(x,y)\otimes l(x,y)=k_c(x,y) \otimes c(x,y)$; what this expression means is that if one averages over a circle at any point in either the low resolution or high resolution calibrated images, the results will agree. It is worth noting that meeting this criterion does not guarantee that the details within any averaging circle will precisely agree with ground truth measurements; empirically however, this simple mathematical approach produces images that appear to be very plausible renditions of the detailed narrow band reflectances of objects in the image. This method does not rely on sampling and model training which offers several advantages over trained models:

\begin{enumerate}
    \item There is no bias between $c(x,y)$ and $l(x,y)$ (by definition), whereas in a trained model, because albedo examples tend to be heavily concentrated around values of 0.2, without careful training data stratification, learned models tend to regress to the mean and hence over estimate dark albedos and underestimate light albedos. 
    \item Because each high resolution image $h(x,y)$ contains the artifacts of changing atmospheric conditions unique to that image, there is no simple model that can be trained to calibrate all high resolution images without incorporating an atmospheric model. Models can however be trained on a per image basis, as described below for the calibration transfer method.
\end{enumerate}

\subsection*{Calibration transfer}
In this machine learning method, a gradient boosted tree model is made by targeting a stratified sample of 10 meter resolution Sentinel-2 derived albedo pixels using as model inputs 10 meter resolution upscaled samples of the  B, G, R, N and panchromatic bands from the high resolution image together with near neighbor pixels and semantic surface category probabilities. The model is trained, tested and validated on the upscaled high resolution image samples, and then applied pixel-wise to the high resolution image. This model relies on the assumption of scale invariance, namely that covariates that predict albedo at 10 meter resolution also represent the albedo at 30 cm resolution. 

\subsection*{Roof segmentation}
In order to find the outline, pitch and azimuth of each roof we rely on the methods described by \cite{batchu2024}. This produces a set of roof segment polygons for which we can compute the median albedo. Because we have information about the surface geometry of each roof segment we are able to improve the accuracy of the albedo estimates by calculating the surface albedo from the image albedo. Sentinel-2 (and by extension our high-resolution albedo inferences) assumes surfaces are both Lambertian and flat. Because of these assumptions, roof segments belonging to the same building with identical roofing materials but with different geometries may have different albedo estimates because of differences in illumination. We calculate the surface albedo, $\alpha_s$, from image albedo, $\alpha_i$, for each of the four albedo estimates using the following equation: 
\begin{equation}
    \left( max\left(D \, \hat{n} \cdot \hat{s}, 0 \right) +I\left(\pi - \gamma \right) / \pi \right) \alpha_s = \left(D \, \hat{k} \cdot \hat{s} +I \right)\alpha_i
\end{equation}

where $D$ is the magnitude of the direct solar irradiance, $\hat{n}$ is the surface normal, $\hat{s}$ is the unit vector pointing toward the solar disc, $I$ is the magnitude of the indirect solar irradiance, $\gamma$ is the zenith angle of the surface, and $\hat{k}$ is the vertical unit vector. We assume a ratio of direct to diffuse radiation of 0.85 to 0.15 and use the solar geometry corresponding to the high resolution imagery (rather than the Sentinel-2 imagery) for correcting the high-resolution inferences (Methods Table 1). Both satellites are sun synchronous for 10:30 am local solar time flyovers. The only solar geometry differences will stem from the obliquity of the PNeo satellite. 

\begin{table}[ht]
\centering
\begin{tabular}{llcc}
\toprule
\textbf{Date} & \textbf{Sensor} & \textbf{Solar Azimuth} & \textbf{Solar Elevation} \\
\midrule
May 5, 2023   & S2    & 142.27 & 64.02 \\
April 30, 2024 & S2   & 149.90 & 62.09 \\
May 8, 2024   & S2    & 145.04 & 61.46 \\
May 16, 2024  & S2    & 143.41 & 63.00 \\
May 16, 2024  & S2    & 143.56 & 63.08 \\
May 5, 2023   & PNeo  & 141.11 & 66.60 \\
May 5, 2023   & PNeo  & 141.21 & 66.49 \\
May 8, 2024   & PNeo  & 144.59 & 63.61 \\
May 16, 2024  & PNeo  & 144.56 & 65.83 \\
May 16, 2024  & PNeo  & 144.37 & 65.79 \\
\bottomrule
\end{tabular}
\caption*{\textbf{Methods Table 1.} Solar azimuth and elevation values for S2 and PNeo sensors for the analysis dates.}
\label{tab:methods1}
\end{table}

\section{Acknowledgements}
Oliver Guinan and Carl Elkin of Google for useful discussions. Shruthi Prabakara for managing our research team. Evan Tachovsky for helping to manage our institutional collaboration and general enthusiasm. 

\section{Author Contributions}
Dave Fork conceived the project and developed the downscaling methodology; Elizabeth Wesley and Dave Fork designed the study and conducted the analysis; Elizabeth Wesley, Dave Fork, and Lucy R. Hutyra led the writing of the manuscript; Salil Banerjee, Vishal Batchu, Aniruddh Chennapragada, Kevin Crossan, Bryce Cronkite-Ratcliff, Jonas Kemp, Yael Mayer, Becca Milman, and Wayne Sun; provided analytical and engineering support; Ellie Delich, Mansi Kansal, Eric Mackres, John C. Platt, Shruthi Prabhakara, Gautam Prasad, Shravya Shetty, and Charlotte Stanton provided program management support; Tristan Goulden provided NEON validation data; All authors contributed to discussion and editing of the manuscript. 

\section{Competing Interests}
The authors declare no competing interests. 

\section{Data Availability}
The Sentinel-2 imagery was generated using publicly available Earth Engine data originating from the European Space Agency. NEON provided validation data from their calibration flight which we have made available in our data archive. High resolution Pléiades Neo (PNeo) imagery was licensed from Airbus for use in Google applications including this research. Very High Resolution optical Pléiades Neo data is available to users for research and development projects from the European Space Agency. Methods described herein could be applied to any RGBN imagery and is encouraged by the authors. All the albedo inferences are available for the roof segments and building footprints at the Harvard Dataverse Data Archive under digital object identifier (DOI) \url{https://doi.org/10.7910/DVN/7T7QDH}. This data is licensed under Creative Commons Attribution 4.0 International (CC BY 4.0). To view a copy of this license, visit \url{https://creativecommons.org/licenses/by/4.0/}

\bibliographystyle{plainnat}   
\bibliography{references}

\end{document}